\documentclass[aps,prresearch,twocolumn,superscriptaddress,nofootinbib,longbibliography,floatfix]{revtex4-2}

\usepackage[T1]{fontenc}
\usepackage[utf8]{inputenc}
\usepackage{amsmath,amssymb,mathtools,bm}
\usepackage{graphicx}
\usepackage{booktabs}
\usepackage{xcolor}
\usepackage{hyperref}

\hypersetup{colorlinks=true,linkcolor=blue,citecolor=blue,urlcolor=blue}

\newcommand{\vac}{\mathrm{vac}}

\newcommand{\sech}{\operatorname{sech}}

\newcommand{\ket}[1]{|#1\rangle}
\newcommand{\bra}[1]{\langle #1|}
\newcommand{\braket}[2]{\langle #1|#2\rangle}

\begin{document}
\title{Squeezing-enhanced Pairwise Fusion of Photonic Qudits}


\author{Pradip Laha}
\email[]{plaha@uni-mainz.de}
\affiliation{Institute of Physics, Johannes Gutenberg-Universit\"at Mainz, Staudingerweg 7, 55128 Mainz, Germany}


\author{Peter van Loock}
\email[]{loock@uni-mainz.de}
\affiliation{Institute of Physics, Johannes Gutenberg-Universit\"at Mainz, Staudingerweg 7, 55128 Mainz, Germany}


\begin{abstract}
Pairwise fusion gates are linear-optical measurements that herald Bell projections onto two-rail subspaces of two \(d\)-rail single-photon qudits. Without ancillary input photons, these passive measurements succeed with probability \(1-d^{-1}\), with all failures confined to the diagonal logical subspace. We show that identical single-mode squeezers applied to the \(2d\) interferometer outputs before photon-number-resolving detection recover part of this structured failure sector. Photon-number parity preserves the successful off-diagonal fusion signatures, while selected all-even patterns yield POVM elements proportional to definite pairwise Bell projectors. We derive the exact logical-space POVM and prove that a diagonal pattern is accepted if and only if its photon-number-imbalance vector has exactly two nonzero components of equal magnitude. The resulting closed elliptic-integral expression increases the ideal success probability, for instance, from \(75\%\) to \(79.62\%\) for \(d=4\), and from \(83.33\%\) to \(87.15\%\) for \(d=6\). With a representative finite detector saturation threshold, \(n_{\rm sat}=7\),
the respective certified values remain \(78.84\%\) and \(86.71\%\). 
These results establish active Gaussian processing as a method for recycling structured measurement failures without ancillary input photons, at the cost of \(2d\) squeezing operations and a larger photon-number range at detection.
\end{abstract}

\maketitle

\section{Introduction}
\label{sec:introduction}
Photonic quantum information processing is measurement-driven. In linear-optical architectures, deterministic photon--photon nonlinearities are generally unavailable, so entangling operations are realized through interference, photon counting, and heralding~\cite{Knill2001,Kok2007}. An ideal Bell-state measurement, or simply Bell measurement (BM), is a joint measurement that projects two quantum systems onto a maximally entangled Bell basis and reports the corresponding Bell label~\cite{Bennett1993,Braunstein1995,BianchiReview2026}. In teleportation, this label specifies the conditional operation on a remote system~\cite{Bennett1993,Bouwmeester1997}; in entanglement swapping, the same mechanism transfers entanglement between systems that have not directly interacted~\cite{Zukowski1993,Pan1998}; and in fusion protocols, a Bell-type entangled projection heralds the connection of initially separate photonic resource states~\cite{Browne2005,GimenoSegovia2015,Bartolucci2023}. BMs and fusion measurements are therefore central operations in linear-optical information processing~\cite{Knill2001,Kok2007}, quantum repeaters and networks~\cite{Briegel1998,Duan2001,Sangouard2011,Azuma2015}, and measurement-based and fusion-based quantum computation~\cite{Browne2005,GimenoSegovia2015,Ewert2016,Bartolucci2023,Paesani2023,Pankovich2024,Schmidt2024Fusion,Song2024}. Because these local measurements are performed repeatedly, even a modest reduction in their failure probability can translate into fewer photons, shorter memory times, shallower fusion networks, and higher entanglement-distribution rates~\cite{Pant2017}. 

The canonical dual-rail two-qubit BM exposes the resource dependence hidden by the abstract definition. Early linear-optical analyzers established that only part of the four-state Bell basis can be identified using two photons and passive optics~\cite{Weinfurter1994,Braunstein1995}. More generally, with vacuum ancillary modes, static passive linear optics, and photon counting, the optimal success probability for unambiguous Bell-state discrimination is \(1/2\) ~\cite{Vaidman1999,Lutkenhaus1999,Calsamiglia2001,ScheelLutkenhaus2004}. This is not a universal limit on photonic BMs, but a theorem for a particular optical resource class, a distinction made explicit by the authors recently~\cite{Laha_2026_HDBSM}. Ancillary photons and enlarged interferometers can exceed the bound~\cite{Grice2011,Ewert2014,Bayerbach2023,Hauser2025,Olivo2018}, while auxiliary entanglement in additional degrees of freedom enables embedded analyzers~\cite{Kwiat1998}. Multiphoton and logical encodings provide another route to high-efficiency and loss-tolerant BMs~\cite{Lee_PRL_2015,Ewert2016,Schmidt_PRA_2019,Hilaire2023}. These encoded measurements also support fault-tolerant and all-optical communication architectures~\cite{Ewert_PRA_2017,Lee_PRA_2019}. Predetection squeezing instead changes the measurement algebra itself ~\cite{ZaidiVanLoock2013,KilmerGuha2019}. These routes trade ideal success against photon number, source complexity, detector requirements, optical depth, and loss tolerance; realistic-imperfection studies show why the ideal success probability alone is insufficient~\cite{Wein2016}.

High-dimensional encoding offers a complementary form of leverage. A single photon distributed over \(d\) modes carries a \(d\)-rail qudit, providing a large logical space without increasing the number of information-carrying photons~\cite{Dada2011,Cozzolino2019,Erhard2020,Wang2020}. On-chip experiments have generated and coherently controlled entangled qudits~\cite{Schaeff2015,Kues2017}, while programmable interferometers and integrated processors support increasingly large mode spaces and high-dimensional optical logical operations~\cite{Clements2016,Imany2019,Zheng2023}. High-dimensional teleportation has been demonstrated experimentally~\cite{Luo2019,Hu2020}, and a recent qutrit entanglement-swapping protocol further illustrates the network-level opportunities and detector tradeoffs of such encodings~\cite{Tanji2026}. Together with multidimensional network proposals and deployed-fiber communication~\cite{Bacco2021,Zahidy2024}, as well as proposals for universal linear-optical qudit computation and deterministic qudit graph-state generation~\cite{Paesani2021,Raissi2024}, these advances make high-dimensional Bell-type measurements and fusion gates increasingly relevant as experimentally motivated operations, rather than formal generalizations of qubit protocols.

The larger Hilbert space does not, however, remove the optical measurement constraint. A complete BM on two \(d\)-level systems must resolve one of the \(d^2\) labels of a generalized Bell basis. In the strict two-photon, vacuum-mode, passive-linear-optical setting, the limitation is stronger than a reduced success probability: for \(d>2\), no conclusive outcome for a full \(d\)-dimensional Bell label exists~\cite{Dusek2001,Calsamiglia2002,Laha_2026_HDBSM}. A useful proposal must therefore specify both its physical resource class and its operational target---complete Bell-label readout, grouped information, or a fusion projection. Recent high-dimensional schemes accordingly introduce populated ancillary states and structured analyzers~\cite{Bharos2025}, active Gaussian processing~\cite{Bianchi2025}, or generalized Type-II fusion constructions~\cite{Ustun2025}. 

Predetection squeezing has, in particular, recently been applied to generalized high-dimensional Bell analyzers, where the active squeezing layer before photon-number-resolving (PNR) detection reshapes photon-counting signatures associated with generalized Bell states~\cite{Bianchi2025}. Our objective here is different. We do not attempt to optimize complete high-dimensional Bell-label discrimination. Instead, we use squeezing as a targeted modification of a fusion measurement whose passive success and failure sectors have a simple, known structure.

Our starting point is the pairwise fusion gate (PFG) introduced by Yamazaki and Azuma~\cite{YamazakiAzuma2025}. Acting on two \(d\)-rail photonic qudits, the ancilla-free PFG projects onto Bell states supported on known two-rail subspaces with success probability \(1-d^{-1}\). Using \(2(2^{k}-1)\) additional high-dimensional ancillary photons, for an integer \(k\geq 1\), the same construction yields PFGs with success probability \(1-d^{-(k+1)}\). Crucially, the ancilla-free failure is not generic: passive interference already converts the entire off-diagonal, or different-rail, logical sector into pairwise Bell projections, whereas all failure probability is confined to the \(d\)-dimensional diagonal, or same-rail, sector. This structure leads to the central question of this work: can an active optical layer act selectively on this diagonal failure sector while leaving the already successful pairwise projections intact?

\begin{figure}
    \centering
    \includegraphics[width=0.99\linewidth]{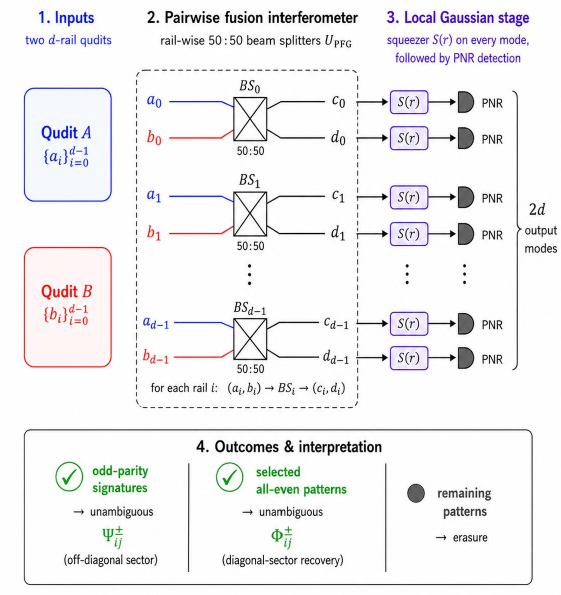}
    \vspace{-4ex}
    \caption{
    Schematic of the local squeezing-enhanced PFG. Two \(d\)-rail photonic qudits enter a rail-wise PFG interferometer, where each matched input pair \((a_i,b_i)\) is mixed on a balanced beam splitter to produce output modes \((c_i,d_i)\). Each of the resulting \(2d\) modes is then locally squeezed by \(S(r)\) before PNR detection. In the lossless setting, odd-parity signatures continue to herald the off-diagonal projections onto \(\ket{\Psi_{ij}^{\pm}}\), while selected all-even patterns enabled by squeezing herald additional diagonal-sector projections onto \(\ket{\Phi_{ij}^{\pm}}\); all remaining outcomes are declared erasures.
    }
    \label{fig:schematic}
\end{figure}

Here, we demonstrate that squeezing can perform precisely this failure-sector engineering. As shown in Fig.~\ref{fig:schematic}, identical single-mode squeezers are appended to all \(2d\) outputs of the ancilla-free PFG before PNR detection.  In the ideal lossless case, squeezing preserves photon-number parity, so the odd-parity signatures that identify the off-diagonal PFG outcomes remain protected. In the all-even diagonal sector, however, squeezing redistributes the photon-number amplitudes and creates additional patterns whose restricted Kraus vectors are proportional to definite pairwise Bell states.



For \(d>2\), the \emph{global} qualification is essential: pairwise diagonal Bell states associated with different rail pairs are not mutually orthogonal because they share diagonal basis states. The accepted probability therefore cannot be obtained by optimizing embedded two-dimensional subspaces separately and adding the results.
Instead, we classify the POVM on the full \(d\)-rail logical space. Every all-even count pattern defines a difference vector whose components are the half-photon-number imbalances between the two output ports of each rail. An outcome is a valid pairwise diagonal Bell projection \emph{if and only if} this vector has exactly two nonzero components of equal magnitude. This criterion gives an exact closed elliptic-integral expression for the ideal-PNR success probability. 


We also develop the lossless finite-resolution theory of the squeezing-enhanced PFG and derive the global criterion for accepted diagonal-sector outcomes.
We find that finite photon-number resolution suppresses, but does not eliminate, the squeezing-induced advantage.

\section{Pairwise fusion with squeezing}
\label{sec:model}

\subsection{Logical space and pairwise Bell projections}

In a \(d\)-rail single-photon encoding, one logical qudit is represented by one photon occupying one of \(d\) orthogonal modes, with all other modes in vacuum. We consider two such photonic qudits, \(A\) and \(B\). Their respective input creation operators are denoted by \(a_i^\dagger\) and \(b_i^\dagger\). The computational basis states are \(\ket{i}_A=a_i^\dagger\ket{\vac}_A\) and \(\ket{j}_B=b_j^\dagger\ket{\vac}_B\), with \(i=0,\ldots,d-1\).
In the two-register Fock space, we write \(\ket{\vac}=\ket{\vac}_A\ket{\vac}_B\) for the joint multimode vacuum.
Since the input contains exactly one photon in each register, the two-qudit logical state lies in the two-photon subspace
\begin{equation}
    \mathcal H_{2\gamma}^{(d)}
    =
    \operatorname{span}
    \left\{
    a_i^\dagger b_j^\dagger\ket{\vac}
    :
    i,j=0,\ldots,d-1
    \right\}.
    \label{eq:logicalSubspaceSecII}
\end{equation}
Equivalently, \(\ket{i}_A\ket{j}_B=a_i^\dagger b_j^\dagger\ket{\vac}\), where \(i,j\in\mathbb{Z}_d\) label the logical single-photon modes of the two qudits, not photon occupation numbers. For each unordered pair of distinct rails \(i<j\), the pairwise Bell states are
\begin{align}
    \ket{\Psi_{ij}^{\pm}}
    &=
    \tfrac{1}{\sqrt2}\Big(
    \ket{i}_A\ket{j}_B
    \pm
    \ket{j}_A\ket{i}_B
    \Big),
    \label{eq:PsiBellSecII}
    \\
    \ket{\Phi_{ij}^{\pm}}
    &=
    \tfrac{1}{\sqrt2}\Big(
    \ket{i}_A\ket{i}_B
    \pm
    \ket{j}_A\ket{j}_B
    \Big).
    \label{eq:PhiBellSecII}
\end{align}
The states \(\ket{\Psi_{ij}^{\pm}}\) belong to the off-diagonal, or different-rail, sector, whereas the states \(\ket{\Phi_{ij}^{\pm}}\) belong to the diagonal, or same-rail, sector. It is useful to introduce the diagonal and off-diagonal projectors
\begin{align}
    \Pi_{\rm diag}
    &=
    \sum_{i=0}^{d-1}
    \ket{i}_A\ket{i}_B
    {}_A\!\bra{i}{}_B\!\bra{i},
    \\
    \Pi_{\rm off}
    &=
    \sum_{\substack{i,j=0\\i\neq j}}^{d-1}
    \ket{i}_A\ket{j}_B
    {}_A\!\bra{i}{}_B\!\bra{j}.    
\end{align}
Their sum
\begin{equation}
    \Pi_{2\gamma}
    =
    \Pi_{\rm off}+\Pi_{\rm diag} =
    \sum_{i,j=0}^{d-1}
    \ket{i}_A\ket{j}_B
    {}_A\!\bra{i}{}_B\!\bra{j},
  \label{eq:pi2gamma}
\end{equation}
is the projector from the full optical Fock space onto the two-photon logical subspace
\(\mathcal H_{2\gamma}^{(d)}\), and hence acts as the identity operator when restricted to that subspace.

There are \(\binom{d}{2}\) unordered rail pairs and two signs for each pair, so each of the two labelled families contains \(d(d-1)\) states. The passive PFG success probability, however, cannot be obtained by counting these labels. The states \(\ket{\Psi_{ij}^{\pm}}\) are mutually orthogonal and form a basis of the \(d(d-1)\)-dimensional off-diagonal sector. By contrast, for \(d>2\), the collection of states \(\ket{\Phi_{ij}^{\pm}}\) is neither mutually orthogonal nor a Bell basis of the diagonal sector. For example, for three distinct rails \(i,j,k\), \(\braket{\Phi_{ij}^{+}}{\Phi_{ik}^{+}} = \frac{1}{2}\).
Different rail pairs therefore share diagonal basis components. A PFG does not implement a complete BM on the full \(d^2\)-dimensional two-qudit space. Rather, each successful outcome heralds a Bell projection supported on a definite two-rail subspace \((i,j)\); all remaining outcomes are treated as erasures.

The passive ancilla-free PFG~\cite{YamazakiAzuma2025} succeeds with unit probability when the input is supported entirely in the off-diagonal sector. Equivalently, its successful outcomes implement restricted Kraus bras proportional to \(\bra{\Psi_{ij}^{\pm}}\) for every \(i<j\), and their POVM elements sum to \(\Pi_{\rm off}\). For an arbitrary logical input state \(\rho\), the passive success probability is therefore
\[
    P_{\rm f}^{\rm(pass)}(\rho)
    =
    \operatorname{Tr}
    \left[
    \Pi_{\rm off}\rho
    \right].
\]
For the maximally mixed logical input,
\(\rho_{\rm mix}=\Pi_{2\gamma}/d^2\), this becomes the off-diagonal fraction of the logical space,
\begin{equation}
    P_{\rm f}(d)
    =
    \frac{d(d-1)}{d^2}
    =
    1-\frac{1}{d}.
    \label{eq:passivePFGSecII}
\end{equation}
Equation~\eqref{eq:passivePFGSecII} is the state-averaged benchmark used throughout this work. The aim of the squeezed construction is to retain unit success conditioned on the off-diagonal sector while recovering a well-defined subset of the diagonal failure sector.

\subsection{Rail-wise PFG interferometer}

The passive interferometer is a product of balanced (50:50) beam splitters (BSs), one for each matched input-rail pair \((a_i,b_i)\). We use the BS phase convention
\begin{align}
    a_i^\dagger
    \longmapsto
    \tfrac{1}{\sqrt2} \left(c_i^\dagger + d_i^\dagger\right),
    \qquad
    b_i^\dagger
    \longmapsto
    \tfrac{1}{\sqrt2} \left(c_i^\dagger -d_i^\dagger\right).
    \label{eq:BSbSecII}
\end{align}
Here \(c_i\) and \(d_i\) denote the two output modes associated with rail \(i\). Other phase conventions differ only by fixed local phases and give the same success probabilities.

The rail-wise interference gives the diagonal and off-diagonal sectors distinct output-parity structures. Each off-diagonal Bell state produces exactly two odd-parity output modes. For \(\ket{\Psi_{ij}^{+}}\), these modes are either \((c_i,c_j)\) or \((d_i,d_j)\), whereas for \(\ket{\Psi_{ij}^{-}}\) they are either \((d_i,c_j)\) or \((c_i,d_j)\). Their locations therefore identify both the rail pair \((i,j)\) and the sign \(\pm\). These parity patterns constitute the passive PFG success outcomes.

By contrast, a diagonal basis input \(\ket{i}_A\ket{i}_B\) is mapped to the two-mode bunched state
\begin{equation}
    \ket{i}_A\ket{i}_B
    \longmapsto
    \tfrac{1}{\sqrt2}\big(\ket{2}_{c_i}\ket{0}_{d_i}
    -
    \ket{0}_{c_i}\ket{2}_{d_i}\big).
    \label{eq:diagonalAfterBSSecII}
\end{equation}
Passive photon counting can identify that the event originated from a same-rail input because the photons remain confined to the output pair \((c_i,d_i)\). However, it also reveals the occupied diagonal rail~\(i\), and therefore implements a same-rail projection proportional to \(\bra{i}_A\bra{i}_B\), rather than erasing the rail information needed to project coherently onto \(\ket{i}_A\ket{i}_B\pm\ket{j}_A\ket{j}_B\), i.e., onto \(\ket{\Phi_{ij}^{\pm}}\). The passive failure probability \(1/d\) is thus not an unstructured erasure: it is confined entirely to the diagonal sector.

\subsection{Local squeezing and effective POVM}

We now append identical single-mode squeezers to all \(2d\) output modes before PNR detection, as shown in Fig.~\ref{fig:schematic}. For a single mode \(c\),
\begin{equation}
    S_c(r)
    =
    \exp\left[
    \tfrac{r}{2}
    \left(
    c^2-c^{\dagger 2}
    \right)
    \right],
    \label{eq:singleModeSqueezerSecII}
\end{equation}
and the full local Gaussian layer is
\begin{equation}
    \mathcal S(r)
    =
    \prod_{i=0}^{d-1}
    S_{c_i}(r)\, S_{d_i}(r).
    \label{eq:multimodeSqueezerSecII}
\end{equation}
A direct spatial implementation therefore uses one squeezing operation on each PFG output mode, for a total of \(2d\) single-mode squeezing operations. Throughout, \(r\) is taken to be real and identical across all modes. Mode-dependent squeezing amplitudes and phases define a broader optimization problem but are not required for the dimension-independent mechanism considered here.

Although the logical input contains exactly two photons, the squeezing layer produces superpositions over higher photon-number sectors. One key property is that it preserves photon-number parity locally. For every output mode \(\mu\),
\begin{equation}
  \left[S_\mu(r),(-1)^{\hat n_\mu}\right]=0,
\end{equation}
where \(\hat n_\mu\) is the photon-number operator of that mode. Hence, in the lossless measurement, the two odd-mode signatures of the passive off-diagonal PFG outcomes remain disjoint from the all-even diagonal sector. Squeezing can therefore redistribute diagonal-sector amplitudes without invalidating the passive PFG outcomes.

The physical readout remains photon counting, while the inserted Gaussian layer changes the POVM induced on the original two-photon logical input space. For a detected photon-number pattern
\begin{equation}
\bm n
=
\left(
n_{c_0},n_{d_0},\ldots,n_{c_{d-1}},n_{d_{d-1}}
\right),
\label{eq:photonPatternSecII}
\end{equation}
the restricted Kraus bra is
\begin{equation}
\bra{\kappa_{\bm n}(r)}
=
\bra{\bm n}
\mathcal S(r)\,
U_{\rm PFG}\,
\Pi_{2\gamma}\,.
\label{eq:krausBraSecII}
\end{equation}
The operators in Eq.~\eqref{eq:krausBraSecII} act from right to left: the logical-subspace projector \(\Pi_{2\gamma}\) defined in Eq.~\eqref{eq:pi2gamma} restricts the input, \(U_{\rm PFG}=\prod_{i=0}^{d-1}U_i^{\rm BS}\) applies the rail-wise interferometer, \(\mathcal S(r)\) applies the local squeezing layer, and \(\bra{\bm n}\) selects the observed photon-counting outcome.

The amplitude for a logical input \(\ket{\psi}\) to produce the detection pattern \(\bm n\) is therefore
\(\braket{\kappa_{\bm n}(r)}{\psi}\), and the corresponding rank-one POVM element on the logical input space is
\begin{equation}
E_{\bm n}(r)
=
\ket{\kappa_{\bm n}(r)}
\bra{\kappa_{\bm n}(r)}.
\label{eq:povmElementSecII}
\end{equation}

A detection pattern is accepted as a successful PFG outcome only when its restricted Kraus bra is nonzero and proportional to one target pairwise Bell bra, i.e.,
\[
\bra{\kappa_{\bm n}(r)}
\propto
\bra{\Psi_{ij}^{\pm}}
\qquad\text{or}\qquad
\bra{\kappa_{\bm n}(r)}
\propto
\bra{\Phi_{ij}^{\pm}}
\]
for a definite rail pair \(i<j\) and a definite sign. All other outcomes are declared erasures; in particular, patterns whose restricted Kraus bras contain contributions from more than one pairwise Bell projection are not counted as PFG successes. This is an operator-level declaration: a successful pattern identifies the pairwise Bell projection implemented by that measurement outcome, rather than discriminating an ensemble of mutually nonorthogonal diagonal states.

With this convention, the accepted photon-counting patterns define the coarse-grained success POVM element
\begin{equation}
M_{\rm suc}(r)
=
\sum_{\bm n\in\mathcal U(r)}
E_{\bm n}(r),
\label{eq:successOperatorSecII}
\end{equation}
where \(\mathcal U(r)\) is the set of accepted photon-number outcomes. Averaging this POVM element over the maximally mixed logical input, \(\rho_{\rm mix}=\Pi_{2\gamma}/d^2\), gives
\begin{equation}
P_{\rm f}(d,r)
=
\operatorname{Tr}
\big[
M_{\rm suc}(r)\rho_{\rm mix}
\big]
=
\frac{1}{d^2}
\operatorname{Tr}_{\mathcal H_{2\gamma}^{(d)}}
\big[
M_{\rm suc}(r)
\big].
\label{eq:successPOVMSecII}
\end{equation}
This averaging convention follows Ref.~\cite{YamazakiAzuma2025}: it weights the actual dimensions of the logical input sectors rather than counting the nonorthogonal labels \(\ket{\Phi_{ij}^{\pm}}\).

For the lossless ideal-PNR measurement, the coarse-grained success element has a simple sector structure. The accepted off-diagonal patterns are exactly the passive PFG success signatures, so the off-diagonal contribution is \(\Pi_{\rm off}\). Squeezing does not change this part of the measurement. Its only effect on the success probability is to make some all-even diagonal-sector patterns implement valid pairwise diagonal Bell projections. These outcomes occur only probabilistically; by rail symmetry, their total accepted probability is the same for every fixed diagonal basis input. Denoting this probability by \(p_{\rm diag}(d,r)\), we obtain
\begin{equation}
M_{\rm suc}(r)
=
\Pi_{\rm off}
+
p_{\rm diag}(d,r)\,
\Pi_{\rm diag},
\label{eq:symmetryReducedPOVMSecII}
\end{equation}
where the exact value of \(p_{\rm diag}(d,r)\) is derived in Sec.~\ref{subsec:closedform}.

Equation~\eqref{eq:symmetryReducedPOVMSecII} also gives the success probability for an arbitrary logical input state \(\rho\):
\begin{equation}
    P_{\rm f}(\rho;d,r)
    =
    \operatorname{Tr}
    \left[
    \Pi_{\rm off}\rho
    \right]
    +
    p_{\rm diag}(d,r)
    \operatorname{Tr}
    \left[
    \Pi_{\rm diag}\rho
    \right].
    \label{eq:generalInputSuccessSecII}
\end{equation}

For the maximally mixed input, these weights are \(1-1/d\) and \(1/d\), respectively. Since \(p_{\rm diag}(d,0)=0\), the passive result is recovered at \(r=0\). 

\section{Exact POVM theory and ideal-PNR performance}
\label{sec:exact}


\subsection{Diagonal sector as a difference-vector measurement}
\label{subsec:diag}

We now classify the accepted all-even diagonal outcomes and evaluate their total probability. Let \(\ket{D_i}\equiv\ket{i}_A\ket{i}_B=a_i^\dagger b_i^\dagger\ket{\vac}\) denote a diagonal computational-basis input. As shown in Eq.~\eqref{eq:diagonalAfterBSSecII}, the rail-wise BS maps \(\ket{D_i}\) to a two-photon bunched superposition in the output modes \(c_i\) and \(d_i\). Accordingly, diagonal inputs contribute only to all-even outcomes.
We write an all-even photon-counting outcome as
\begin{equation}
    n_{c_i}=2q_i^c,
    \qquad
    n_{d_i}=2q_i^d,
    \qquad
    q_i^c,q_i^d\in\mathbb N_0 .
    \label{eq:evenPattern}
\end{equation}

For a single mode, define the squeezed-vacuum and squeezed-two-photon amplitudes
\begin{equation}
    v_q(r)=\bra{2q}S(r)\ket{0},
    \qquad
    u_q(r)=\bra{2q}S(r)\ket{2}.
    \label{eq:vuDef}
\end{equation}
The required matrix elements are given in Appendix~\ref{app:squeezed-matrix-elements}. For \(r>0\), the ratio \(u_q(r)/v_q(r)\) is affine in \(q\):
\begin{equation}
    R_q(r)
    \equiv
    \frac{u_q(r)}{v_q(r)}
    =
    \frac{\tanh r}{\sqrt2}
    -
    \frac{\sqrt2\,q}{\sinh r\cosh r}.
    \label{eq:Rq}
\end{equation}
The passive endpoint \(r=0\) is evaluated directly from the original amplitudes, or equivalently by taking the continuous limit of the final probability expressions; the ratio in Eq.~\eqref{eq:Rq} is not evaluated at \(r=0\).

For an all-even pattern
\begin{equation}
    \bm q
    =
    \left(
    q_0^c,q_0^d,\ldots,q_{d-1}^c,q_{d-1}^d
    \right),    
\end{equation}
the Kraus bra restricted to
\(\mathcal D=\operatorname{span}\{\ket{D_i}\}_{i=0}^{d-1}\)
is
\begin{equation}
    \bra{K_{\bm q}}
    =
    \sum_{i=0}^{d-1}
    A_i(\bm q)\bra{D_i},
    \label{eq:Kqgeneral}
\end{equation}
where \(A_i(\bm q)=\braket{K_{\bm q}}{D_i}\). Since \(\ket{D_i}\) enters the squeezing layer as the bunched superposition of Eq.~\eqref{eq:diagonalAfterBSSecII}, while all rails \(\ell\ne i\) are initially in vacuum, \(A_i(\bm q)\) factorizes as
\begin{equation}
    A_i(\bm q)
    =
    \frac{1}{\sqrt2}
    \left[
    u_{q_i^c}\,v_{q_i^d}
    -
    v_{q_i^c}\,u_{q_i^d}
    \right]
    \prod_{\ell\ne i}
    v_{q_\ell^c}\,v_{q_\ell^d}.
    \label{eq:Aigeneral}
\end{equation}
For compactness, the argument \(r\) of \(u_q\) and \(v_q\) is suppressed in intermediate expressions. The relative minus sign is inherited from the BS phase convention in Eq.~\eqref{eq:diagonalAfterBSSecII}. Using \(u_q=R_qv_q\), the bracket in Eq.~\eqref{eq:Aigeneral} becomes
\[
    v_{q_i^c}v_{q_i^d}
    \left[
    R_{q_i^c}-R_{q_i^d}
    \right]
    =
    -
    \frac{\sqrt2\left(q_i^c-q_i^d\right)}
    {\sinh r\cosh r}
    v_{q_i^c}v_{q_i^d}.
\]
Therefore, for \(r>0\),
\(A_i(\bm q)=C_{\bm q}(r)\, m_i\), where
\begin{equation}
    m_i=q_i^c-q_i^d
    =
    \tfrac{1}{2}
    \big(n_{c_i}-n_{d_i}\big)
    \label{eq:mi}
\end{equation}
is the half-photon-number imbalance between the two output ports of rail \(i\), and
\begin{equation}
    C_{\bm q}(r)
    =
    -
    \frac{1}{\sinh r\cosh r}
    \prod_{\ell=0}^{d-1}
    v_{q_\ell^c}(r)v_{q_\ell^d}(r)
    \label{eq:Cq}
\end{equation}
is a common amplitude prefactor for the pattern \(\bm q\). Equation~\eqref{eq:Kqgeneral} therefore becomes
\begin{equation}
    \bra{K_{\bm q}}
    =
    C_{\bm q}(r)
    \sum_{i=0}^{d-1}
    m_i\,\bra{D_i}.
    \label{eq:differencevector}
\end{equation}
Although the separated prefactor in Eq.~\eqref{eq:Cq} is not defined at \(r=0\), the physical amplitudes in Eq.~\eqref{eq:Aigeneral} have a regular passive limit. All statements based on Eqs.~\eqref{eq:Rq}--\eqref{eq:differencevector} are therefore understood for \(r>0\), with \(r=0\) treated by continuity at the probability level.

Each all-even squeezed detection outcome is thus represented in the diagonal subspace by the integer rail-imbalance vector
\begin{equation}
    \bm m=(m_0,\ldots,m_{d-1})\in\mathbb Z^d . 
    \label{eq:m_def}
\end{equation}
The prefactor \(C_{\bm q}(r)\) determines the probability weight, whereas the direction of \(\bm m\) determines the projected diagonal state. In particular, \(\bm m=\bm 0\) gives a vanishing Kraus bra on the diagonal logical subspace.

Let \(\bm e_i\) denote the \(i\)-th standard basis vector in \(\mathbb Z^d\), and let \(h=1,2,\ldots\) be the common magnitude of the nonzero imbalances. A nonzero all-even outcome implements a pairwise diagonal Bell projection if and only if
\begin{equation}
    \bm m
    =
    \pm h\bm e_i
    \pm h\bm e_j,
    \qquad
    i<j .
    \label{eq:Bellcriterion}
\end{equation}
Equation~\eqref{eq:Bellcriterion} is exactly the requirement that the restricted Kraus bra have support on only two diagonal basis states, \(\bra{D_i}\) and \(\bra{D_j}\), with coefficients of equal magnitude. Equal signs give a bra proportional to \(\bra{D_i}+\bra{D_j}\), and hence to \(\bra{\Phi_{ij}^{+}}\) while opposite signs give a bra proportional to \(\bra{D_i}-\bra{D_j}\), and hence to \(\bra{\Phi_{ij}^{-}}\), up to an ignorable overall phase. The criterion is independent of the squeezing strength: \(r\) changes the probability assigned to each pattern but not whether its imbalance vector represents a target pairwise Bell projection. The necessity and sufficiency of Eq.~\eqref{eq:Bellcriterion} are proved in Appendix~\ref{app:difference-vector-proof}. 

All other all-even outcomes are rejected by the PFG success rule. A single nonzero component gives a Kraus bra proportional to \(\bra{D_i}\) for one rail \(i\); two nonzero components of unequal magnitude give a nonmaximally weighted two-rail superposition; and support on more than two rails gives a higher-dimensional diagonal superposition. Such outcomes may be relevant for other measurements, but they are not accepted as pairwise fusion successes here.




\subsection{Exact ideal-PNR success probability}
\label{subsec:closedform}


We now sum the probabilities of all all-even outcomes satisfying \eqref{eq:differencevector}.  Let
\begin{equation}
w_q(r)
=
|v_q(r)|^2
=
\sech r\,
\frac{(2q)!}{2^{2q}(q!)^2}
\tanh^{2q}r
\label{eq:wq}
\end{equation}
be the probability of detecting \(2q\) photons from a squeezed-vacuum input. Consider a fixed diagonal input \(\ket{D_i}\). For an all-even pattern \(\bm q\), Eqs.~\eqref{eq:Cq} and \eqref{eq:differencevector} give
\begin{equation}
\left|
\braket{K_{\bm q}}{D_i}
\right|^2
=
\frac{m_i^2}
{\sinh^2 r\cosh^2 r}
\prod_{\ell=0}^{d-1}
w_{q_\ell^c}(r)w_{q_\ell^d}(r).
\label{eq:PatternProb}
\end{equation}
The diagonal recovery probability is obtained by summing Eq.~\eqref{eq:PatternProb} over all accepted imbalance patterns. For each partner rail \(j\neq i\), imbalance magnitude \(h\geq1\), the accepted patterns satisfy
\[
m_i=\pm h,
\quad
m_j=\pm h,
\quad
m_\ell=0
\quad
(\ell\neq i,j).
\]
Thus,
\begin{align}
p_{\rm diag}(d,r)
&=
\sum_{j\neq i}
\sum_{h=1}^{\infty}
\sum_{\substack{
\bm q:\,
|m_i|=h,\\
|m_j|=h,\\
m_\ell=0;(\ell\neq i,j)
}}
\left|
\braket{K_{\bm q}}{D_i}
\right|^2 .
\label{eq:PdiagRawSum}
\end{align}

This sum factorizes rail by rail. To evaluate it, define the squeezed-vacuum autocorrelations
\begin{equation}
Z_h(r)
=
\sum_{q=0}^{\infty}
w_{q+h}(r)\,w_q(r),
\qquad
h=0,1,2,\ldots .
\label{eq:ZhDef}
\end{equation}
For \(h\geq1\), \(Z_h(r)\) is the weight associated with one specified signed imbalance between two squeezed-vacuum modes; summing over the two signs gives \(2Z_h(r)\). The zero-imbalance weight is \(Z_0(r)\).

For the occupied rail \(i\), the factor \(m_i^2\) in Eq.~\eqref{eq:PatternProb} contributes \(h^2\), and summing over the two signs of \(m_i\) gives
\[
\frac{2h^2Z_h(r)}
{\sinh^2 r\cosh^2 r}.
\]
For the partner rail \(j\), summing over the two signs of \(m_j\) gives \(2Z_h(r)\). Each of the remaining \(d-2\) spectator rails has zero imbalance and contributes \(Z_0(r)\). Therefore,
\begin{align}
p_{\rm diag}(d,r)
&=
\sum_{j\neq i}
\sum_{h=1}^{\infty}
\left[
\frac{2h^2Z_h(r)}
{\sinh^2 r\cosh^2 r}
\right]\!\!
\big[
2Z_h(r)
\big]
Z_0(r)^{d-2}
\nonumber\\
&=
(d-1)
\frac{4Z_0(r)^{d-2}}
{\sinh^2 r\cosh^2 r}
\sum_{h=1}^{\infty}
h^2Z_h(r)^2 .
\label{eq:PdiagSeries}
\end{align}

Introducing \(y=\tanh^2(2r)\) with $0\leqslant y<1$, the above equation reduces exactly to
\begin{equation}
    p_{\rm diag}(d,r)
    =
    \frac{d-1}{4}\,
    y\,
    (1-y)^{(d-1)/2}
    \left[
    \frac{2\,\mathrm K(y)}{\pi}
    \right]^{d-2}\,.
    \label{eq:pdiagMainSim}
\end{equation}
Here, \(\mathrm K(y) = \int_0^{\pi/2} d\theta/\sqrt{1-y\sin^2\theta}\)
is the complete elliptic integral of the first kind in the parameter convention. The \emph{reduction} to Eq.~\eqref{eq:pdiagMainSim} and the absolute convergence of Eq.~\eqref{eq:PdiagSeries} are established in Appendix~\ref{app:diagonal-success}.

Equation~\eqref{eq:pdiagMainSim} is the exact ideal-PNR probability that a fixed diagonal basis input produces an accepted pairwise Bell projection. By rail-permutation symmetry, it is independent of the occupied rail and is the diagonal-sector coefficient of the success POVM introduced in Sec.~\ref{sec:model}. 

The full state-averaged success probability is obtained by adding the unchanged off-diagonal contribution and weighting the diagonal contribution by the diagonal fraction \(1/d\):
\begin{equation}
    P_{\rm f}^{\rm id}(d,r)
    =
    1-\frac{1}{d}
    +
    \frac{1}{d}\,
    p_{\rm diag}(d,r)\,.
    \label{eq:totalSuccessSecIII}
\end{equation}
At \(r=0\), \(y=0\) and hence \(p_{\rm diag}(d,0)=0\), so Eq.~\eqref{eq:totalSuccessSecIII} reduces to the passive value \(1-1/d\). For every finite \(r>0\), the expression in Eq.~\eqref{eq:pdiagMainSim} is strictly positive. 

\begin{figure*}[t]
    \centering
    \includegraphics[width=\textwidth]{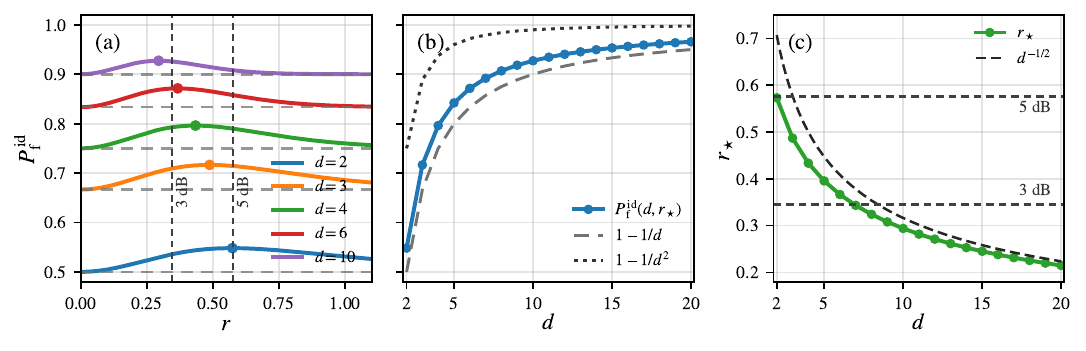}
    \vspace{-5ex}
\caption{
Ideal-PNR performance of the squeezing-enhanced PFG.
(a) Success probability \(P_{\rm f}^{\rm id}(d,r)\) versus the uniform squeezing parameter \(r\) for representative dimensions. Dashed horizontal lines show the passive ancilla-free PFG values \(1-1/d\), and markers denote the optimized operating points.
(b) Optimized success probability \(P_{\rm f}^{\rm id}(d,r_\star)\) for \(d=2,\ldots,20\), compared with the passive benchmark \(1-1/d\) and the lossless intrinsic \(k=1\) ancilla-assisted benchmark \(1-1/d^2\).
(c) Optimal squeezing parameter \(r_\star\) versus \(d\). The dashed reference curve \(d^{-1/2}\) shows the analytically predicted large-\(d\) scaling.
}
\label{fig:ideal-performance}
\end{figure*}

\subsection{Optimization and physical scaling}
\label{subsec:ideal-numerics}

Note that \(y\to1\) as \(r\to\infty\) and therefore \(y\) never reaches 1 at finite \(r\). For optimization over nonzero finite squeezing, we have \(0< y<1\).
Figure~\ref{fig:ideal-performance} shows the ideal-PNR performance obtained from Eq.~\eqref{eq:totalSuccessSecIII}. The diagonal contribution vanishes at \(r=0\) and again in the limit \(r\rightarrow\infty\), and therefore reaches a maximum at a finite squeezing strength. Physically, squeezing must generate equal-magnitude imbalances in the occupied rail and one partner rail while leaving all remaining rails with zero imbalance. 

Since the off-diagonal contribution in
Eq.~\eqref{eq:totalSuccessSecIII} is independent of \(r\), maximizing the total success probability is equivalent to maximizing \(p_{\rm diag}(d,r)\). Writing
\(
    y_\star=\tanh^2(2r_\star),
\)
the unique finite-squeezing optimum is determined by
\begin{equation}
    (d-2)
    \frac{\mathrm E(y_\star)}
         {\mathrm K(y_\star)}
    =
     d-4+3y_\star\,,
    \label{eq:exactOptimalityCondition}
\end{equation}
where \(\mathrm E(y) = \int_0^{\pi/2} \sqrt{1-y\sin^2\theta}\,d\theta\)
is the complete elliptic integral of the second kind. The derivation of Eq.~\eqref{eq:exactOptimalityCondition}, the uniqueness of its solution, and its asymptotic expansion are given in Appendix~\ref{app:diagonal-success}.

The optimized success probability \(P_{\rm f}^{\rm id}(d,r_\star)\), with \( r_\star = \frac{1}{2} \operatorname{arctanh}\sqrt{y_\star}\), is displayed in Fig.~\ref{fig:ideal-performance}(b). It exceeds the passive value \(1-1/d\) for every dimension considered. The absolute improvement is largest at small and moderate dimensions. In particular, the success probability increases from \(0.75\) to \(0.7962\) for \(d=4\), and from \(0.8333\) to \(0.8715\) for \(d=6\). 

At large \(d\), expansion of Eq.~\eqref{eq:exactOptimalityCondition} gives
\begin{align}
    y_\star
    &=
    \frac{4}{d}
    +
    O(d^{-2}),
    \\
    r_\star^2
    &=
    \frac{1}{d}
    +
    O(d^{-2}),\\
    p_{\rm diag}(d,r_\star)
    &=
    e^{-1}
    +
    O(d^{-1}).
    \label{eq:PdiagAsymptotic}
\end{align}
Thus, the optimized measurement asymptotically recovers a fraction \(e^{-1}\) of the diagonal failure sector. Since this sector has weight \(1/d\) in the maximally mixed logical input, the absolute gain over the passive PFG is
\[
    \Delta P_{\rm f}^{\rm id}(d)
    \equiv
    P_{\rm f}^{\rm id}(d,r_\star)
    -
    \left(1-\frac{1}{d}\right)
    =
    \frac{1}{ed}
    +
    O(d^{-2}).
\]
The recovered fraction therefore remains finite with increasing dimension, while the absolute improvement decreases because the diagonal failure sector itself occupies only a fraction \(1/d\) of the logical space.

For \(d=2\), Eq.~\eqref{eq:exactOptimalityCondition} gives the exact solution
\( y_\star=\frac{2}{3}\),
\( r_\star = \frac{1}{2} \operatorname{arctanh}\sqrt{\frac{2}{3}},\)
and hence
\[
    P_{\rm f}^{\rm id}(2,r_\star)
    =
    \frac{1}{2}
    +
    \frac{1}{12\sqrt3}
    =
    0.548113\ldots .
\]
This value is the \(d=2\) endpoint of the dimension-independent difference-vector rule, rather than an optimization of the specialized complete qubit Bell analyzer of Ref.~\cite{ZaidiVanLoock2013}.

\section{Finite detector resolution}
\label{sec:finite}

The ideal-PNR analysis of Sec.~\ref{sec:exact} assumes lossless detection with unbounded photon-number resolution. We now keep the measurement lossless but impose a finite detector saturation threshold \(n_{\rm sat}\): photon numbers \(0,1,\ldots,n_{\rm sat}-1\) are resolved, while the outcome \(\nu=n_{\rm sat}\) represents the saturation bin \(n\ge n_{\rm sat}\). Because finite resolution does not change the parity of resolved counts, the odd-parity off-diagonal signatures remain disjoint from the all-even diagonal sector within the resolved window. We therefore use a conservative certified rule: any event containing a saturated detector is declared an erasure, and every resolved event is accepted only if it satisfies the same pairwise imbalance criterion as in the ideal-PNR analysis.

For the lossless case, saturation discards high-count events without altering resolved counts. The largest resolved even and odd half-photon indices are
\begin{equation}
    Q_e
    =
    \left\lfloor
    \tfrac{1}{2} \left(n_{\rm sat}-1\right)
    \right\rfloor,
    \qquad
    Q_o
    =
    \left\lfloor
    \tfrac{1}{2}\left(n_{\rm sat}-2\right)
    \right\rfloor ,
    \label{eq:QeQo}
\end{equation}
where \(\lfloor\cdot\rfloor\) is the floor function. Thus,
\(2q<n_{\rm sat}\) for \(q\leqslant Q_e\), while
\(2q+1<n_{\rm sat}\) for \(q\leqslant Q_o\).


For an off-diagonal PFG outcome to remain within the trusted finite window, the two odd modes must lie inside the resolved odd window, while the remaining \(2d-2\) even modes must lie inside the resolved even window. The resulting probability is
\begin{equation}
    p_{\rm off}^{(n_{\rm sat})}(d,r)
    =
    \left[
    \sum_{q=0}^{Q_o}
    o_q(r)
    \right]^2
    \left[
    \sum_{q=0}^{Q_e}
    w_q(r)
    \right]^{2d-2}.
    \label{eq:finiteOffdiag}
\end{equation}
Here,  for \(q=0,1,2,\ldots\),
\begin{equation}
    o_q(r)
    = 
    \left|
    \bra{2q+1}S(r)\ket{1}
    \right|^2 =
    \sech^3 r\,
    \frac{(2q+1)!}{2^{2q} \,(q!)^2}
    \tanh^{2q}r,
    \label{eq:oddWeightsFinite}
\end{equation}
is the probability for a squeezed single-photon, as derived in Appendix~\ref{app:squeezed-matrix-elements}.

\begin{figure}[t]
    \centering
    \includegraphics[width=\columnwidth]{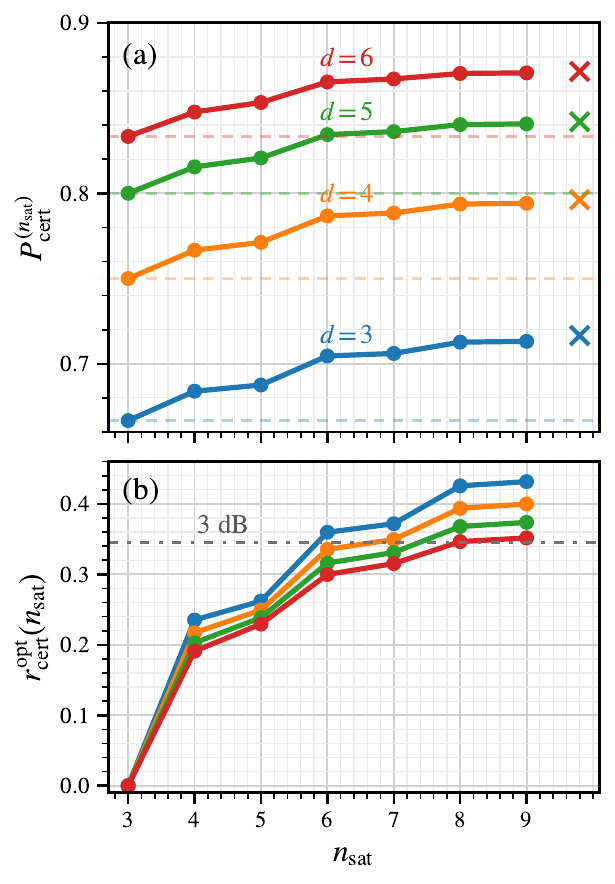}
    \vspace{-5ex}
\caption{
Lossless finite-resolution performance of the squeezing-enhanced PFG under the certified zero-error declaration rule.
(a) Optimized probability \(P_{\rm cert}^{(n_{\rm sat})}\) as a function of the detector saturation threshold \(n_{\rm sat}\), shown for \(d=3,4,5,6\). 
Dashed horizontal lines mark the passive ancilla-free PFG benchmarks \(1-1/d\), and crosses at the right show the corresponding optimized ideal-PNR values.
(b) Optimized squeezing parameter \(r_{\rm cert}^{\rm opt}\) for the same dimensions. The step structure results from the finite detector window: successive changes in \(n_{\rm sat}\) alternately enlarge the resolved odd and even photon-number sectors, whereas the additional diagonal recovery depends on the all-even sector.
}
    \label{fig:finite-resolution-usd}
\end{figure}

The diagonal-sector recovery retains the difference-vector structure derived in Sec.~\ref{subsec:diag}, except that the squeezed-vacuum autocorrelations are restricted to the resolved even window:
\begin{equation}
    Z_h^{(Q_e)}(r)
    =
    \sum_{q=0}^{Q_e-h}
    w_{q+h}(r)w_q(r),
    \qquad
    h=0,\ldots,Q_e .
    \label{eq:truncatedZ}
\end{equation}
The certified diagonal Bell-recovery probability is therefore
\begin{equation}
    p_{\rm diag}^{(n_{\rm sat})}(d,r)
    =
    \frac{
    4(d-1)
    \left[
    Z_0^{(Q_e)}(r)
    \right]^{d-2}
    }{
    \sinh^2 r\cosh^2 r
    }
    \sum_{h=1}^{Q_e}
    h^2
    \left[
    Z_h^{(Q_e)}(r)
    \right]^2 .
    \label{eq:finiteDiag}
\end{equation}
As in the ideal-PNR expression, the value at \(r=0\) is understood through its continuous limit.


Combining the off-diagonal and diagonal contributions gives
\begin{equation}
    P_{\rm cert}^{(n_{\rm sat})}(d,r)
    =
    \left(
    1-\frac{1}{d}
    \right)
    p_{\rm off}^{(n_{\rm sat})}(d,r)
    +
    \frac{1}{d}
    p_{\rm diag}^{(n_{\rm sat})}(d,r).
    \label{eq:finiteCertTotal}
\end{equation}
This quantity is optimized over \(r\) in Fig.~\ref{fig:finite-resolution-usd}. Because every pattern containing a saturated bin is declared an erasure, Eq.~\eqref{eq:finiteCertTotal} is a conservative certified success probability rather than an estimate based on an assumed model for unresolved counts.

Figure~\ref{fig:finite-resolution-usd} shows that finite photon-number resolution suppresses, but does not eliminate, the squeezing-induced advantage. For \(n_{\rm sat}=3\), the trusted window is too narrow to certify a gain. The additional diagonal contribution becomes visible at \(n_{\rm sat}=5\), and by \(n_{\rm sat}=7\) the optimized probabilities are already close to their ideal-PNR values. The \(n_{\rm sat}=7\) results are summarized in Table~\ref{tab:finite-resolution-main}. This threshold is chosen as a representative finite-resolution setting motivated by photon-number-resolving capability demonstrated in transition-edge sensors and emerging SNSPD-based architectures~\cite{Lita2008,Fukuda2011,Morais2024,Schapeler2024}. Detector efficiency, count rate, and scalable multimode operation remain separate experimental requirements.  

\begin{table}[t]
\caption{
Certified finite-window performance for saturation threshold \(n_{\rm sat}=7\). The detector resolves photon numbers \(0,\ldots,6\), while the outcome \(7\) denotes saturation; any pattern containing a saturated bin is declared an erasure. The column \(P_{\rm f}(d)=1-1/d\) gives the passive ancilla-free PFG value, while \(P_{\rm f}^{\rm id}(d,r_\star)\) gives the optimized ideal-PNR value from Sec.~\ref{sec:exact}. The improvement \(\Delta P\) is measured relative to \(P_{\rm f}(d)\), and the squeezing in decibels is \(20r/\ln 10\).
}
\label{tab:finite-resolution-main}
\begin{ruledtabular}
\begin{tabular}{ccccccc}
\(d\) &
\(P_{\rm f}(d)\) &
\(r_{\rm cert}^{\rm opt}\) &
dB &
\(P_{\rm cert}^{(7)}\) &
\(P_{\rm f}^{\rm id}(d,r_\star)\) &
\(\Delta P\) \\
\hline
3 & 0.6667 & 0.3718 & 3.2300 & 0.7061 & 0.7166 & 0.0394 \\
4 & 0.7500 & 0.3495 & 3.0350 & 0.7884 & 0.7962 & 0.0384 \\
5 & 0.8000 & 0.3310 & 2.8750 & 0.8362 & 0.8420 & 0.0362 \\
6 & 0.8333 & 0.3153 & 2.7380 & 0.8671 & 0.8715 & 0.0337
\end{tabular}
\end{ruledtabular}
\end{table}

The optimized squeezing decreases with increasing \(d\), consistently with the ideal-PNR scaling in Sec.~\ref{subsec:ideal-numerics}. Finite resolution shifts the optimum toward smaller \(r\) as stronger squeezing transfers more probability into photon-number components outside the trusted window.

\section{Conclusion and outlook}
\label{sec:conclusion}

We have shown that local single-mode squeezing can enhance an ancilla-free PFG for \(d\)-rail photonic qudits without changing the encoding or supplying ancillary input photons. The mechanism uses the structure of the passive PFG: the off-diagonal sector is already converted into pairwise Bell projections, while all passive failures lie in the diagonal sector. In the lossless setting, the squeezing layer leaves the successful off-diagonal signatures intact and converts a certified subset of all-even diagonal outcomes into additional pairwise Bell projections. The accepted diagonal outcomes are characterized by a global rail-imbalance criterion: an all-even pattern is accepted if and only if its imbalance vector has exactly two nonzero components of equal magnitude. This avoids overcounting among the nonorthogonal diagonal pairwise states and yields an exact success POVM with a closed elliptic-integral expression for the recovered diagonal contribution.

The resulting improvement is moderate but systematic. In the ideal-PNR limit, the success probability increases from \(75\%\) to \(79.62\%\) for \(d=4\), and from \(83.33\%\) to \(87.15\%\) for \(d=6\). At large \(d\), the optimized measurement recovers a fraction \(e^{-1}\) of the diagonal failure sector, giving an absolute gain \(1/(ed)+O(d^{-2})\) over the passive PFG. Finite photon-number resolution alone does not remove the advantage: for saturation threshold \(n_{\rm sat}=7\), resolving photon numbers \(0,\ldots,6\), the certified lossless probabilities remain \(78.84\%\) for \(d=4\) and \(86.71\%\) for \(d=6\). These gains require \(2d\) squeezing operations, phase stability, low insertion loss, and increased PNR dynamic range.

Several extensions are natural. More general Gaussian layers, including passive multimode mixing, rail-dependent squeezing, and optimized phases, may recover a larger part of the structured failure space. The all-even outcomes rejected by the present pairwise criterion may also be useful, since many correspond to known higher-dimensional diagonal superpositions rather than featureless failures. Hybrid squeezing--ancilla PFGs provide another possible route, but their advantage should be established by a PFG-specific optimization rather than by resource counting alone. Beyond the lossless model considered here, realistic imperfections remain crucial: post-squeezer loss can change photon-number parity and mix signatures that are disjoint in the ideal analysis. Quantifying the effects of detector inefficiency, insertion loss, nonuniform squeezing, phase drift, beam-splitter imbalance, mode mismatch, PNR errors, and possible approximate declaration rules is therefore an important next step toward assessing experimental feasibility.

More broadly, this work illustrates a strategy of failure-space engineering: preserve the part of a measurement that already succeeds, and act selectively on the structured support of its failure POVM. Here, high-dimensional encoding supplies a large passively successful sector, while local Gaussian processing converts a rigorously characterized part of the remaining failure space into certified entangling outcomes. The same idea may be useful for other photonic measurements whose failures occupy an identifiable subspace that can be selectively reshaped by accessible optical operations.

\begin{acknowledgments}
We thank P. A. Ameen Yasir for discussions. We acknowledge funding from the BMFTR in Germany for support via PhotonQ, QR.N, QuKuK, QuaPhySI, and also from the EU project CLUSTEC (Grant Agreement No. 101080173).
\end{acknowledgments}




\appendix

\section{Single-mode squeezed matrix elements}
\label{app:squeezed-matrix-elements}

This appendix gives the single-mode matrix elements used in Sec.~\ref{sec:exact}. We use the real squeezing convention
\begin{equation}
    S(r)
    =
    \exp\left[
    \frac{r}{2}
    \left(
    a^2-a^{\dagger 2}
    \right)
    \right],
    \qquad
    r\ge0 .
    \label{eq:appSqueezerDef}
\end{equation}
With this convention,
\begin{align}
    S(r)
    =
    \exp\left[
    -\frac{\tanh r}{2}a^{\dagger 2}
    \right]
    &\exp\left[
    -\ln(\cosh r)
    \left(
    a^\dagger a+\frac12
    \right)
    \right]\nonumber\\
    &\times\exp\left[
    \frac{\tanh r}{2}a^2
    \right].
    \label{eq:appNormalOrdered}
\end{align}

The squeezed-vacuum amplitudes are
\begin{align}
    v_q(r)
    \equiv
    \bra{2q}S(r)\ket{0}
    =
    \sqrt{\sech r}\,
    \frac{\sqrt{(2q)!}}{2^q q!}
    \left[-\tanh r\right]^q ,
    \label{eq:appVq}
\end{align}
with \(q=0,1,2,\ldots\). Taking the modulus square gives
\begin{align}
    w_q(r)
    &\equiv
    |v_q(r)|^2
    =
    \sech r\,
    \frac{(2q)!}{2^{2q}(q!)^2}
    \tanh^{2q}r .
    \label{eq:appWqFactorial}
\end{align}

We also need the two-photon input matrix element 
\begin{equation}
    u_q(r)
    \equiv
    \bra{2q}S(r)\ket{2}.
    \label{eq:appUqDef}
\end{equation}
The rightmost exponential in Eq.~\eqref{eq:appNormalOrdered} gives
\begin{equation}
    \exp\left[
    \frac{\tanh r}{2}a^2
    \right]\ket{2}
    =
    \ket{2}
    +
    \frac{\tanh r}{\sqrt2}\ket{0}.
    \label{eq:appTwoPhotonStep}
\end{equation}
Substituting this into Eq.~\eqref{eq:appNormalOrdered} and projecting onto \(\ket{2q}\) gives
\begin{equation}
    u_q(r)
    =
    v_q(r)
    \left[
    \frac{\tanh r}{\sqrt2}
    -
    \frac{\sqrt2 q}{\sinh r\cosh r}
    \right].
    \label{eq:appUq}
\end{equation}
Therefore
\begin{equation}
    R_q(r)
    \equiv
    \frac{u_q(r)}{v_q(r)}
    =
    \frac{\tanh r}{\sqrt2}
    -
    \frac{\sqrt2 q}{\sinh r\cosh r}.
    \label{eq:appRq}
\end{equation}

For the finite-resolution analysis in Sec.~\ref{sec:finite}, we also need the odd-photon distribution obtained by squeezing a single photon. Defining
\begin{equation}
    o_q(r)
    =
    \left|
    \bra{2q+1}S(r)\ket{1}
    \right|^2 ,
    \qquad
    q=0,1,2,\ldots ,
    \label{eq:appOqDef}
\end{equation}
one finds
\begin{align}
    \bra{2q+1}S(r)\ket{1}
    &=
    \sech^{3/2} r\,
    \frac{\sqrt{(2q+1)!}}{2^q q!}
    \left[-\tanh r\right]^q ,
    \nonumber\\
    o_q(r)
    &=
    \sech^3 r\,
    \frac{(2q+1)!}{2^{2q}(q!)^2}
    \tanh^{2q}r .
    \label{eq:appOq}
\end{align}

\section{Proof of the difference-vector criterion}
\label{app:difference-vector-proof}

This appendix proves the difference criterion used in Sec.~\ref{subsec:diag}. We work only in the diagonal subspace
\begin{equation}
    \mathcal D
    =
    \operatorname{span}
    \left\{
    \ket{D_i}\equiv\ket{i}_A\ket{i}_B
    \right\}_{i=0}^{d-1}.
    \label{eq:appDspace}
\end{equation}
For an all-even detection pattern, the main-text derivation gives the restricted Kraus bra
\begin{equation}
    \bra{K_{\bm q}}
    =
    C_{\bm q}(r)
    \sum_{\ell=0}^{d-1}
    m_\ell \bra{D_\ell},
    \qquad
    m_\ell=q_\ell^c-q_\ell^d .
    \label{eq:appKdifference}
\end{equation}
The scalar \(C_{\bm q}(r)\) fixes the probability scale but not the normalized state selected inside \(\mathcal D\). Thus the relevant object for Bell-projectability is the integer rail-imbalance vector \(\bm m=(m_0,\ldots,m_{d-1})\).

A valid pairwise diagonal Bell projection requires \(\bra{K_{\bm q}}\) to be proportional to
\[
    \bra{\Phi_{ij}^{\pm}}
    =
    \frac{
    \bra{D_i}\pm\bra{D_j}
    }{\sqrt2},
    \qquad i<j .
\]
Equivalently, there must exist a nonzero scalar \(\lambda\), a rail pair \(i<j\), and a sign \(s=\pm1\) such that
\begin{equation}
    C_{\bm q}(r)m_\ell
    =
    \lambda
    \left(
    \delta_{\ell i}
    +
    s\,\delta_{\ell j}
    \right)
    \quad
    \text{for all } \ell .
    \label{eq:appCoeffCompare}
\end{equation}
Because the same scalar \(C_{\bm q}(r)\) multiplies every component, Eq.~\eqref{eq:appCoeffCompare} can hold only if \(\bm m\) has exactly two nonzero components, on rails \(i\) and \(j\), and those two components have equal magnitude. Thus
\begin{equation}
    \bm m
    =
    h\bm e_i+s h\bm e_j,
    \quad
    i<j,
    \quad
    h\in\mathbb Z\setminus\{0\},
    \quad
    s=\pm1 .
    \label{eq:appCriterionSigned}
\end{equation}
Conversely, if Eq.~\eqref{eq:appCriterionSigned} holds, then
\begin{equation}
    \bra{K_{\bm q}}
    =
    h C_{\bm q}(r)
    \left(
    \bra{D_i}
    +
    s\bra{D_j}
    \right),
    \label{eq:appCriterionSufficiency}
\end{equation}
which is proportional to \(\bra{\Phi_{ij}^{+}}\) for \(s=+1\) and to \(\bra{\Phi_{ij}^{-}}\) for \(s=-1\), up to fixed local phases. This proves necessity and sufficiency.

Absorbing the overall sign of \(h\) into the irrelevant proportionality constant, the accepted rail-imbalance vectors are therefore
\begin{equation}
    \bm m
    =
    \pm h\bm e_i
    \pm h\bm e_j,
    \qquad
    i<j,
    \qquad
    h=1,2,\ldots .
    \label{eq:appBellcriterion}
\end{equation}
Equal signs select the \(+\) Bell sign and opposite signs select the \(-\) Bell sign. All other all-even patterns are rejected by the PFG success rule.

\section{Evaluation and optimization of the diagonal success probability}
\label{app:diagonal-success}

Starting from the autocorrelation representation in Eq.~\eqref{eq:PdiagSeries}, we first establish its absolute convergence and then reduce it in terms of complete elliptic integrals. Throughout this appendix, we use
\[
    x=\tanh^2 r,
    \qquad
    0\leqslant x<1.
\]
Since
\[
    \frac{(2q)!}{2^{2q}(q!)^2}
    =
    \frac{1}{4^q}\binom{2q}{q}
    \leqslant slant 1,
\]
and \(\sech r = \sqrt{1-x}\), Eq.~\eqref{eq:wq} implies
\[
    w_q(r)
    \leqslant 
    \sqrt{1-x}\,x^q .
\]
Therefore,
\begin{align}
    Z_h(r)
    &=
    \sum_{q=0}^{\infty}
    w_{q+h}\, w_q 
    \leqslant 
    (1-x)
    \sum_{q=0}^{\infty}
    x^{q+h}x^q 
    =
    \frac{x^h}{1+x}\nonumber.    
\end{align}
Consequently,
\[
    \sum_{h=1}^{\infty}h^2Z_h(r)^2
    \leqslant 
    \frac{1}{(1+x)^2}
    \sum_{h=1}^{\infty}h^2x^{2h}
\]
Now, for \(|x|<1\), it can be shown that
\[
\sum_{h=1}^{\infty}h^2x^{2h} = \frac{x^2(1+x^2)} {(1-x^2)^3}. 
\]
Substituting this, we obtain 
\[
    \sum_{h=1}^{\infty}h^2Z_h(r)^2
    \leqslant 
    \frac{x^2(1+x^2)}
    {(1+x)^2(1-x^2)^3}
    <
    \infty
\]
for \(0\leqslant x<1\), equivalently for every finite \(r\). This bound is used only to establish absolute convergence and to justify the Fourier manipulations below; the exact value of the weighted autocorrelation sum is derived in Eq.~\eqref{eq:appWeightedZClosed}.

\subsection{Reduction to an elliptic-integral expression}

With \(x=\tanh^2 r\), the squeezed-vacuum probabilities can be written as
\begin{equation}
    w_q(r)
    =
    \sqrt{1-x}\,
    \frac{1}{4^q}
    \binom{2q}{q}
    x^q .
    \label{eq:appWqX}
\end{equation}
Using the binomial series identity
\[
    \frac{1}{\sqrt{1-u}}
    =
    \sum_{q=0}^{\infty}
    \frac{1}{4^q}
    \binom{2q}{q}
    u^q,
    \qquad |u|<1,
\]
the generating function of the weights is
\begin{equation}
    W_x(z)
    \equiv
    \sum_{q=0}^{\infty}w_q(r)z^q
    =
    \sqrt{\frac{1-x}{1-xz}}.
    \label{eq:appGeneratingFunction}
\end{equation}
On the unit circle, define
\begin{align}
    F_x(\theta)
    &=
    W_x(e^{i\theta})W_x(e^{-i\theta})
    =
    \frac{1-x}
    {\sqrt{1-2x\cos\theta+x^2}}.
    \label{eq:appFx}
\end{align}
On the other hand, expanding the product of generating functions gives
\begin{align}
    F_x(\theta)
    = \sum_{m,n=0}^\infty w_m(r) w_n(r) e^{i (m-n)\theta}.
    \label{eq:appFxTheta}
\end{align}
Now group terms by the difference \(h=m-n\). For \(h=0\), 
\[ \sum_{n=0}^\infty w_n^2(r) = Z_0(r).\]
For \(h>0\), set \(m=n+h\). Then the \(+h\) terms give 
\[\sum_{n=0}^\infty w_{n+h}(r) w_n(r) e^{ih\theta} = Z_h(r) e^{ih\theta},\]
and similarly the \(-h\) terms give \( Z_h(r) e^{-ih\theta}\). Therefore,
\begin{align}
    F_x(\theta) = Z_0(r)
    +
    2\sum_{h=1}^{\infty}
    Z_h(r)\cos(h\theta),
    \label{eq:appZFourierSeries}
\end{align}
so the autocorrelations \(Z_h(r)\) are the Fourier coefficients of
\(F_x(\theta)\).

The zero-imbalance coefficient is
\begin{align}
    Z_0(r)
    &=
    (1-x)
    \sum_{q=0}^{\infty}
    \left[
    \frac{1}{4^q}
    \binom{2q}{q}
    \right]^2
    x^{2q}
    \nonumber\\
    &=
    \frac{2(1-x)}{\pi}
    \mathrm K(x^2),
    \label{eq:appZ0Elliptic}
\end{align}
where we used the standard series representation of the complete elliptic integral of the first kind,
\begin{equation}
    \mathrm K(m)
    =
    \int_0^{\pi/2}
    \frac{d\theta}
    {\sqrt{1-m\sin^2\theta}}
    =
    \frac{\pi}{2}
    \sum_{q=0}^{\infty}
    \left[
    \frac{1}{4^q}
    \binom{2q}{q}
    \right]^2
    m^q .
    \label{eq:appKDefinition}
\end{equation}
Here \(m\) denotes the elliptic-integral parameter.

The weighted autocorrelation sum can be evaluated using Parseval's
identity. Differentiating Eq.~\eqref{eq:appZFourierSeries} gives
\[
    F_x'(\theta)
    =
    -2
    \sum_{h=1}^{\infty}
    h\, Z_h(r)\sin(h\theta),
\]
and hence
\begin{equation}
    \sum_{h=1}^{\infty}
    h^2Z_h(r)^2
    =
    \frac{1}{4\pi}
    \int_0^{2\pi}
    \left|
    F_x'(\theta)
    \right|^2
    d\theta.
    \label{eq:appParseval}
\end{equation}
From Eq.~\eqref{eq:appFx},
\[
    F_x'(\theta)
    =
    -
    \frac{
    x(1-x)\sin\theta
    }{
    \left(
    1-2x\cos\theta+x^2
    \right)^{3/2}
    }.
\]
Therefore,
\begin{align}
    \sum_{h=1}^{\infty}
    h^2Z_h(r)^2
    &=
    \frac{x^2(1-x)^2}{4\pi}
    \int_0^{2\pi}
    \!\!
    \frac{\sin^2\theta\,d\theta}
    {
    \left(
    1-2x\cos\theta+x^2
    \right)^3
    }.
    \label{eq:appWeightedZIntegral}
\end{align}

For completeness, introduce
\[
    J(a,b)
    =
    \int_0^{2\pi}
    \frac{d\theta}{a-b\cos\theta}
    =
    \frac{2\pi}{\sqrt{a^2-b^2}},
    \qquad
    a>|b|.
\]
Since
\[
    \int_0^{2\pi}
    \frac{\sin^2\theta\,d\theta}
    {(a-b\cos\theta)^3}
    =
    \frac12
    \left(
    \frac{\partial^2J}{\partial a^2}
    -
    \frac{\partial^2J}{\partial b^2}
    \right),
\]
setting \(a=1+x^2\) and \(b=2x\) gives
\begin{equation}
    \int_0^{2\pi}
    \frac{\sin^2\theta\,d\theta}
    {
    \left(
    1-2x\cos\theta+x^2
    \right)^3
    }
    =
    \frac{\pi}{(1-x^2)^3}.
    \label{eq:appTrigonometricIntegral}
\end{equation}
Substitution into Eq.~\eqref{eq:appWeightedZIntegral} yields
\begin{equation}
    \sum_{h=1}^{\infty}
    h^2Z_h(r)^2
    =
    \frac{x^2}
    {4(1-x)(1+x)^3}.
    \label{eq:appWeightedZClosed}
\end{equation}

Using
\[
    \sinh^2r\cosh^2r
    =
    \frac{x}{(1-x)^2},
\]
Eqs.~\eqref{eq:PdiagSeries},
\eqref{eq:appZ0Elliptic}, and
\eqref{eq:appWeightedZClosed} give
\begin{equation}
    p_{\rm diag}(d,r)
    =
    (d-1)
    \frac{x(1-x)}{(1+x)^3}
    \left[
    \frac{2(1-x)}{\pi}
    \mathrm K(x^2)
    \right]^{d-2}.
    \label{eq:appPdiagX}
\end{equation}

We now introduce
\begin{equation}
    y
    =
    \frac{4x}{(1+x)^2}
    =
    \tanh^2(2r).
    \label{eq:appYDefinition}
\end{equation}
The quadratic transformation
\[
    \mathrm K(x^2)
    =
    \frac{1}{1+x}
    \mathrm K
    \left(
    \frac{4x}{(1+x)^2}
    \right)
\]
and the identity
\[
    1-y
    =
    \left(
    \frac{1-x}{1+x}
    \right)^2
\]
reduce Eq.~\eqref{eq:appPdiagX} to
\begin{equation}
    p_{\rm diag}(d,r)
    =
    \frac{d-1}{4}\,
    y(1-y)^{(d-1)/2}
    \left[
    \frac{2\mathrm K(y)}{\pi}
    \right]^{d-2},
    \label{eq:appPdiagCompact}
\end{equation}
which is Eq.~\eqref{eq:pdiagMainSim} of the main text. 

At \(r=0\), one has \(y=0\) and hence
\(p_{\rm diag}(d,0)=0\). For every finite \(r>0\),
\(0<y<1\), so Eq.~\eqref{eq:appPdiagCompact} is strictly positive.
In the opposite limit \(r\rightarrow\infty\), \(y\rightarrow1\), and
the algebraic factor \((1-y)^{(d-1)/2}\) dominates the logarithmic
divergence of \(\mathrm K(y)^{d-2}\), giving
\(p_{\rm diag}(d,r)\rightarrow0\).

\subsection{Exact optimization condition}

Because \(y=\tanh^2(2r)\) is strictly increasing for \(r\geq0\),
maximizing \(p_{\rm diag}(d,r)\) over \(r\) is equivalent to maximizing
Eq.~\eqref{eq:appPdiagCompact} over \(0<y<1\). Its logarithmic
derivative is
\begin{equation}
    \frac{d}{dy}
    \ln p_{\rm diag}
    =
    \frac{1}{y}
    -
    \frac{d-1}{2(1-y)}
    +
    (d-2)
    \frac{\mathrm K'(y)}
    {\mathrm K(y)}.
    \label{eq:appLogDerivative}
\end{equation}
Introduce the complete elliptic integral of the second kind,
\begin{equation}
    \mathrm E(y)
    =
    \int_0^{\pi/2}
    \sqrt{1-y\sin^2\theta}\,d\theta.
    \label{eq:appEDefinition}
\end{equation}
Using
\begin{equation}
    \mathrm K'(y)
    =
    \frac{
    \mathrm E(y)-(1-y)\mathrm K(y)
    }{
    2y(1-y)
    },
    \label{eq:appKDerivative}
\end{equation}
the stationary condition
\(d\ln p_{\rm diag}/dy=0\) becomes
\begin{equation}
    (d-2)
    \frac{\mathrm E(y_\star)}
         {\mathrm K(y_\star)}
    =
    d-4+3y_\star,
    \quad
    r_\star
    =
    \frac12
    \operatorname{arctanh}\sqrt{y_\star},
    \label{eq:appOptimalityCondition}
\end{equation}
which is Eq.~\eqref{eq:exactOptimalityCondition} of the main text.

The solution of Eq.~\eqref{eq:appOptimalityCondition} is unique. To
see this, define
\begin{align}
    R(y)
    &=
    \frac{\mathrm E(y)}{\mathrm K(y)},
\\
    G_d(y)
    &=
    (d-2)R(y)-d+4-3y.    
\end{align}
Differentiating \(R(y)\) gives
\begin{equation}
    R'(y)
    =
    -
    \frac{
    \left[
    R(y)-(1-y)
    \right]^2
    +
    y(1-y)
    }{
    2y(1-y)
    }
    <0
    \label{eq:appRDerivative}
\end{equation}
for \(0<y<1\). Hence
\[
    G_d'(y)
    =
    (d-2)R'(y)-3
    <0.
\]
Moreover,
\[
    \lim_{y\to0^+}G_d(y)=2,
    \qquad
    \lim_{y\to1^-}G_d(y)=1-d<0.
\]
Thus, \(G_d(y)\) has exactly one zero in \((0,1)\). Since
\(p_{\rm diag}\) vanishes at both endpoints and is positive in the interval, this zero gives the unique global maximum.

\subsubsection*{Example cases}
For \(d=2\), Eq.~\eqref{eq:appOptimalityCondition} reduces to
\[
    -2+3y_\star=0,
\]
so
\begin{equation}
    y_\star
    =
    \frac23,
    \qquad
    r_\star
    =
    \frac12
    \operatorname{arctanh}\sqrt{\frac23}.
    \label{eq:appD2Optimum}
\end{equation}
Equation~\eqref{eq:appPdiagCompact} then gives the exact analytical expression 
\[
    p_{\rm diag}(2,r_\star)
    =
    \frac{1}{6\sqrt3}.
\]
Thus
\begin{equation}
    P_{\rm f}^{\rm id}(2,r_\star)
    =
    \frac12
    +
    \frac{1}{12\sqrt3}
    =
    0.5481125\ldots .
    \label{eq:appD2Success}
\end{equation}


For \(d=4\), Eq.~\eqref{eq:appPdiagCompact} becomes
\begin{equation}
    p_{\rm diag}(4,r)
    =
    \frac{3}{\pi^2}\,
    y(1-y)^{3/2}\,
    \mathrm K^2(y).
    \label{eq:appD4Pdiag}
\end{equation}
The stationarity condition in Eq.~\eqref{eq:appOptimalityCondition}
reduces to \(    2\,\frac{\mathrm E(y_\star)}{\mathrm K(y_\star)} = 3y_\star\),
or equivalently \(2\mathrm E(y_\star)=3y_\star\mathrm K(y_\star)\).
This equation has a unique solution in \(0<y_\star<1\), which we compute numerically. We obtain
\begin{equation*}
    y_\star = 0.489762\ldots,
    \quad
    r_\star = \frac12\operatorname{arctanh}\sqrt{y_\star} = 0.433483 .
\end{equation*}
At this point,
\begin{align}
    p_{\rm diag}(4,r_\star) &=  0.184796, \nonumber\\
    P_{\rm f}^{\rm id}(4,r_\star)
    &=
    \frac{3}{4}
    +
    \frac{1}{4}p_{\rm diag}(4,r_\star)
    =    0.796199.
\end{align}

\subsection{Large-\texorpdfstring{\(d\)}{d} expansion}

For small \(y\),
\begin{equation}
    \frac{\mathrm E(y)}{\mathrm K(y)}
    =
    1
    -
    \frac{y}{2}
    -
    \frac{y^2}{16}
    -
    \frac{y^3}{32}
    +
    O(y^4).
    \label{eq:appERatioExpansion}
\end{equation}
Substituting the ansatz
\[
    y_\star
    =
    \frac{a_1}{d}
    +
    \frac{a_2}{d^2}
    +
    \frac{a_3}{d^3}
    +
    O(d^{-4})
\]
into Eq.~\eqref{eq:appOptimalityCondition} and matching successive
powers of \(1/d\) gives
\begin{equation}
    y_\star
    =
    \frac{4}{d}
    -
    \frac{18}{d^2}
    +
    \frac{90}{d^3}
    +
    O(d^{-4}).
    \label{eq:appYStarExpansion}
\end{equation}
Since 
\(
    r_\star
    =
    \frac12
    \operatorname{arctanh}\sqrt{y_\star}
\),
this implies
\begin{equation}
    r_\star^2
    =
    \frac{1}{d}
    -
    \frac{11}{6d^2}
    +
    O(d^{-3}).
    \label{eq:appRStarExpansion}
\end{equation}

Finally, using
\[
    \ln
    \left[
    \frac{2\mathrm K(y)}{\pi}
    \right]
    =
    \frac{y}{4}
    +
    \frac{7y^2}{64}
    +
    O(y^3)
\]
in Eq.~\eqref{eq:appPdiagCompact} gives
\begin{equation}
    p_{\rm diag}(d,r_\star)
    =
    e^{-1}
    \left[
    1-\frac{13}{4d}
    +
    O(d^{-2})
    \right].
    \label{eq:appPdiagOptExpansion}
\end{equation}
The corresponding gain over the passive PFG is
\begin{equation*}
    \Delta P_{\rm f}^{\rm id}(d)
    =
    \frac{p_{\rm diag}(d,r_\star)}{d}
    =
    \frac{1}{ed}
    \left[
    1-\frac{13}{4d}
    +
    O(d^{-2})
    \right].
    \label{eq:appGainExpansion}
\end{equation*}

\section{Numerical details for the ideal-PNR optimization}
\label{app:ideal-numerical-data}

This appendix records the optimized ideal-PNR data used in Fig.~\ref{fig:ideal-performance}. The diagonal-sector contribution was evaluated from the closed elliptic-integral expression in Eq.~\eqref{eq:pdiagMainSim}, and the total success probability from Eq.~\eqref{eq:totalSuccessSecIII}. For each dimension, the optimum was obtained by solving the stationarity condition in Eq.~\eqref{eq:exactOptimalityCondition} for
\(y_\star=\tanh^2(2r_\star)\), followed by
\[
    r_\star
    =
    \frac12\operatorname{arctanh}\sqrt{y_\star}.
\]
The squeezing in decibels is quoted as \(20r_\star/\ln 10\).

For reproducibility, the root of Eq.~\eqref{eq:exactOptimalityCondition} was found by bracketed one-dimensional root finding on \(0<y<1\). The endpoint values \(p_{\rm diag}(d,0)=0\) and \(\lim_{r\to\infty}p_{\rm diag}(d,r)=0\) were checked explicitly, so the unique interior stationary point is the global maximum. The residual in Eq.~\eqref{eq:exactOptimalityCondition} was below \(10^{-13}\). 
As an independent check, direct maximization of Eq.~\eqref{eq:totalSuccessSecIII} over \(r\) reproduced the same optimized probabilities to the displayed precision. The optimized values satisfy
\[
    P_{\rm f}(d)
    \leqslant 
    P_{\rm f}^{\rm id}(d,r_\star)
    \leqslant 
    1,
\]
and their large-\(d\) behavior agrees with the asymptotic expansion derived in Appendix~\ref{app:diagonal-success}.

\bibliography{references}

\end{document}